\documentclass[traditabstract]{aa} 
\usepackage{graphicx}
\usepackage{txfonts}
\usepackage{natbib}
\bibpunct{(}{)}{;}{a}{}{,} 
\newcommand{\threej}[6]
{ \left( \begin{array}{ccc}
#1&#2&#3\\
#4&#5&#6
\end{array} \right) }

\newcommand{\sixj}[6]
{ \left\{ \begin{array}{ccc}
#1&#2&#3\\
#4&#5&#6
\end{array} \right\} }

\begin{document}
   \title{An Open Source, Massively Parallel Code for Non-LTE
     Synthesis and Inversion of Spectral Lines and Zeeman-induced
     Stokes Profiles}

   \author{H. Socas-Navarro\inst{1, 2}, J. de la Cruz Rodr\'\i
     guez\inst{3}, A. Asensio Ramos\inst{1, 2}, J. Trujillo Bueno\inst{1, 2,4}, B. Ruiz Cobo\inst{1, 2}}

   \institute{Instituto de Astrof\'\i sica de Canarias,
     Avda V\'\i a L\'actea S/N, La Laguna 38205, Tenerife, Spain
   \and
   Departamento de Astrof\'\i sica, Universidad de La Laguna, 38205, 
   La Laguna, Tenerife, Spain 
   \and
   Institute for Solar Physics, Dept. of Astronomy, Stockholm University, 
     Albanova University Center, 10691 Stockholm, Sweden
   \and 
   Consejo Superior de Investigaciones Científicas, Spain
 }

   \date{\today}

   \authorrunning{Socas-Navarro et al}
   \titlerunning{A Code for NLTE Stokes Synthesis and Inversion}

\newcommand {\FeI} {\ion{Fe}{i}}
\newcommand {\ScII} {\ion{Sc}{ii}}
\newcommand {\NiI} {\ion{Ni}{i}}
\newcommand {\OI} {\ion{O}{i}}
\newcommand {\CaII} {\ion{Ca}{ii}}
\newcommand {\ltau} {$\log(\tau_{5000})$}

\abstract{With the advent of a new generation of solar telescopes and
  instrumentation, the interpretation of chromospheric observations
  (in particular, spectro-polarimetry) requires new, suitable
  diagnostic tools. This paper describes a new code, NICOLE, that has
  been designed for Stokes non-LTE radiative transfer, both for
  synthesis and inversion of spectral lines and Zeeman-induced
  polarization profiles, spanning a wide range of atmospheric heights,
  from the photosphere to the chromosphere. The code fosters a number
  of unique features and capabilities and has been built from scratch
  with a powerful parallelization scheme that makes it suitable for
  application on massive datasets using large supercomputers. The
  source code is being publicly released, with the idea of
  facilitating future branching by other groups to augment its
  capabilities.  }


   \maketitle
%

\section{Introduction}

The relevance of chromospheric line diagnostics has increased
dramatically in the last decade among solar scientists. This is due to
(or evidenced by, depending on how one looks at it) very significant
advances in both numerical simulations and spectro-polarimetric
observations. The largest instrumental projects for the next decades,
namely the DKIST (formerly known as ATST, \citealt{KRW+11};
\citealt{KRK+03}), the EST (\citealt{CBC+13}; \citealt{EST10}) and
Solar-C (\citealt{KWH+12}; \citealt{STH+11}) have all been designed
with chromospheric magnetometry as a top priority. Moreover, several
important modern facilities such as SOLIS (\citealt{BCG+13};
\citealt{PSH+11}), Gregor (\citealt{DLP+12}; \citealt{SVvdL+12}), the
SST with CRISP (\citealt{2003SPIE.4853..341S};
\citealt{2008ApJ...689L..69S}), the DST and the Big Bear NST
(\citealt{CGN13}; \citealt{GC12}), foster remarkable
chromospheric-observing capabilities.

Numerical simulations of the solar atmosphere have grown notably in
size, scope and complexity (\citealt{S12}; \citealt{RS11}). A
particularly noteworthy effort in this context is the development of
numerical MHD simulations of the magnetic chromosphere
(\citealt{KCD+14,GCH+11}). The simulated magnetic structures are still
of relatively low field strengths, but this limitation is of technical
nature. Hopefully, more processing power and new developments in
numerical methods will permit higher flux densities in the near
future.

Bridging the gap between the new ground-breaking observations and
simulations requires complex modeling and diagnostic tools. NICOLE is
a step in this direction. Capable of Non-LTE (hereafter NLTE) spectral
line calculations, it is suitable for the analysis of chromospheric
lines and their polarization profiles in the Zeeman regime. The user
is able to synthesize spectral profiles from large simulation
datacubes, allowing a direct comparison with observations (it is
possible to include the instrumental profile in the calculation).
Conversely, the code inversion engine is able to work on the observed
spectral data to infer relevant atmospheric parameters (such as
temperatures, magnetic field vector or Doppler velocities) which may
provide interesting information or be compared directly with the
simulations. Other existing NLTE codes that share some (but not all)
of NICOLE's features are HAZEL (\citealt{ARTBLdI08}), RH
(\citealt{U01}) and Porta (\citealt{STB13}).

NICOLE has been designed from the beginning to work on massive
datasets, e.g. large simulation snapshots or high-resolution
observations. The code implements a simple but efficient master-slave
scheme using the widely available MPI (Message-Passing Interface)
parallelization. This design makes it suitable for any architecture,
including the most powerful supercomputers with over a thousand
processors. With its 1.5D approach (meaning that each model column is
treated as a horizontally-infinite atmosphere), almost ideal
paralellization is achieved even for the largest number of processors.

The discussion presented in this section has been thus far focused on
solar physics only but this tool is of great potential usefulness in
other areas of astrophysics, as well. The code can easily provide
flux-calibrated spectra of late-type stars. The capability for
inversion of stellar spectra has been implemented following the work
of \citeauthor{APRCGL98} (\citeyear{APRCGL98}; see also
\citealt{APGLL+00,APBA+01}) and works similarly to their code
MISS. Chemical abundances may be inverted using NICOLE, as well, which
might be an interesting capability for studies of solar/stellar
compositions.

NICOLE has been released to the community as an open source project
under the GPL license\footnote{http://www.gnu.org/copyleft/gpl.html}, which
means that it may be copied, altered and redistributed, as long as any
resulting product is also distributed openly to the community. Users
are welcome, and in fact encouraged, to branch out their own version
of NICOLE to improve it, augment it or to implement new features. The
source code is currently hosted at the following repository:
https://github.com/hsocasnavarro/NICOLE

\section{Code description}

Although NICOLE has been almost entirely written from scratch and
incorporates many novel modules and elements, it builds upon previous
experience with other very popular radiative transfer codes. The
structure of the inversion mode in NICOLE is similar to that of
SIR (Stokes Inversion based on Response-functions,
\citealt{RCdTI92}). In the NLTE module, the structure and variable
naming is similar to MULTI (\citealt{SC85}). The NLTE iterative core
is an implementation of the method described in
\citet{SNTB97}. The inversion module works in the same way as the
code of \citet{SNRCTB98,SNTBRC00a}. The following is a list of the
approximations and limitations that have driven the design of NICOLE:

\begin{itemize}
\item Statistical equilibrium: The NLTE atomic populations are
  computed assuming instantaneous balance between all transitions
  going into and out of each atomic level. Effects such as
  time-dependent ionization are thus neglected in the synthesis
  (although it could have been previously incorporated in the
  computation of the model atmosphere in the synthesis mode).
  The tests presented in \citet{dlCRSNC+12} support the validity of
  this assumption in a realistic scenario involving the inversion of
  \CaII \ lines. Furthermore, \citet{LCRvdV12} point out that it is
  also a suitable strategy for H$\alpha$ synthesis, provided that the
  MHD model accounts for such time-dependent ionization. Nevertheless,
  there might be other situations in which this approximation would be
  less adequate.
\item Complete angle and frequency redistribution (CRD): This
  approximation states that the frequency and direction of an emitted
  photon is independent of the frequency and direction of a previously
  absorbed one by the atomic system. \citet{U89}
  demonstrated that this approximation works very well for the \CaII
  \, infrared triplet lines. Other lines, such as \CaII \, H and~K
  exhibit some significant discrepancies near the core (but not at the core
  itself), between CRD and full computations.
\item Polarization induced by the Zeeman effect: NICOLE does not
  account for polarization produced by scattering processes or
  modified by the Hanle effect. It is therefore more suitable for
  application on Stokes~$I$ and~$V$, and for all the Stokes profiles
  only when the magnetic field is strong enough (typically in active
  regions). Observing away from the solar limb also helps to reduce
  the possible influence of scattering and Hanle (de-)polarization on
  the linear polarization profiles.
\item Field-free NLTE populations: The statistical equilibrium
  equations are solved neglecting the presence of a magnetic
  field. This is usually a good approximation since the lines are
  often much broader than the Zeeman splitting (\citealt{R69}).
\item Hydrostatic equilibrium: This approximation is employed {\em
  only} in inversion mode and {\em only} for the computation of the
  density scale. It affects mostly the conversion of optical to
  geometrical depth and, to some extent, the background
  opacities. Otherwise, strong line profiles are usually rather
  insensitive to density and pressure changes.
\item Blends: Spectral calculations (both syntheses and inversions)
  may include an arbitrary number of lines with the only limitation
  that all the NLTE lines must be of the same element. Line blends are
  treated consistently in the final formal solution, including their
  polarization profiles. However, the NLTE atomic level populations
  are computed without considering blends.
\item Collisional damping: The code incorporates the classic Unsold
  formula (\citealt{U55}) and the more recent formalism of
  \citet{AO95} and \citet{BAOM98}.
\item 1.5D calculation: Although the code works with three dimensional
  datacubes, each column is treated independently, as if it were
  infinite in the horizontal direction. This approximation works well
  in LTE and when computing strong NLTE lines. The reason for the
  latter is that in the line core the opacity is so high that the
  photons have a short mean free path. Therefore, the populations are
  controlled by the environmental conditions in their immediate
  surroundings. A more quantitative assessment of this approximation
  has been presented in \citet{dlCRSNC+12}.
\item Hyperfine structure: Lines with hyperfine structure may be
  seamlessly integrated in the spectral synthesis or inversion, simply
  by supplying the appropriate atomic data in the configuration
  file. However, this mode usually has a significant performance
  impact (e.g., \citealt{SN14}).
\item Flexible node location: The inversion nodes for NICOLE may be
  specified manually by the user and do not need to be equispaced. This
  enables a more efficient distribution of nodes through the
  atmosphere, packing them more densely where more information is
  available and spreading them out in areas where the observations are
  less sensitive.
\item Bezier-interpolant formal solvers: NICOLE implements a number of
  options for the formal solution method. A very interesting new
  routine is based on Bezier interpolation and makes the code more
  robust and stable. More details are provided in
  Section~\ref{sec:formalsolvers} below.
\end{itemize}

\section{The equation of state}

The equation of state (EoS) establishes one or more constraints that
relate the various fundamental parameters defining the state of the
plasma.  Solving the EoS to determine physical variables, such as
electron pressure, internal energy or the H$^-$ negative ion density,
to name a few examples, is often a necessary intermediate step in a
broad range of numerical codes for radiative transfer or MHD
simulation. It is not trivial to solve the EoS when partial ionization
and molecule formation are considered. 

For the purposes of the calculations involved in NICOLE, the plasma
state is defined by its temperature ($T$), gas density ($\rho$), gas
pressure ($P_g$), electron pressure ($P_e$) and the following number
densities, needed for the background opacities: neutral hydrogen atoms
(H), protons (H$^+$), negative hydrogen ions (H$^-$), hydrogen
molecules (H$_2$) and ionized hydrogen molecules (H$^+_2$). All of
these parameters may be supplied as input if desired. Alternatively,
one could supply two of them (temperature plus one of density, gas
pressure or electron pressure) and NICOLE will use the EoS to solve
for the rest. If the option to impose hydrostatic equilibrium is set,
then only the temperature stratification and the upper boundary
condition for the electron pressure are needed. This is actually how
the code works in inversion mode.

In NICOLE the solution of the EoS is divided in two, generally
independent, steps. The first step computes the distribution of the
various H populations (H, H$^+$, H$^-$, H$_2$, H$^+_2$). The second
step deals with the relationship among the thermodynamical parameters
($T$, $\rho$, $P_g$ and $P_e$).  In both cases there are three
different methods to solve the EoS in NICOLE that the user may choose
from.

\subsection{Full ICE solution}

In order to determine the density of particles in the solar plasma we
need to solve not only the atomic ionization system but also the
chemical equilibrium among all the possible molecules. NICOLE
implements the instantaneous chemical equilibrium (ICE) calculation of
\citet{ARTBC+03, ARSN05}, with a compilation of data for a total of 273
molecules. Obviously, dealing with such a large number of molecules
results in a very demanding computation at each gridpoint, having to
solve a non-linear system of 273 equations and unknowns. In addition to
the computing time demanded by this approach, there is also the
problem that sometimes the iteration fails to converge. Only a very
small percentage of the points suffer this convergence issue but
nevertheless it might still be problematic for some applications in
which stability over a large number of calculations is a strict
requirement. Therefore, two other options, faster and more stable,
have been implemented as well.

\subsection{Restricted ICE solution}

This is basically the same as the full solution except that only two H
molecules are considered, H$_2$ and H$^+_2$. The calculation is then
faster and more stable than the full ICE but it might still fail
(albeit rarely) and is slower than the NICOLE option (see below).

\subsection{The NICOLE method}

We have developed a new procedure that avoids iteration and is
therefore perfectly stable and faster than the previously discussed
options. It is based on the realization that one needs to know only
how much of the H is in molecular form to derive all other relevant
parameters in a straightforward manner. We trained an artificial
neural network (ANN) using a large database of ($T$, $P_e$, $m$)
values for which we had previously solved the full ICE system with the
273 molecules. The third input parameter $m$ characterizes the plasma
metallicity (in a logarithmic scale), so that all elements heavier than
$Z$=2 have their abundances scaled by this factor. The ANN and the
algorithms are very similar to those described in \citet{SN03} and
\citet{SN05f}. The training set was initially computed starting from a
uniform distribution of $T$ (between 1,500 and 10,000~K), $m$ (between
-1.5 and 0.5) and $log(P_g)$ (between -3 and 6 with $P_g$ in
dyn$~$cm$^{-2}$). We used this initial dataset to study some
properties of the distribution but the actual ANN training was done
using a more optimal set, as explained below.

\begin{figure*}
  \centering
  \includegraphics[width=\textwidth]{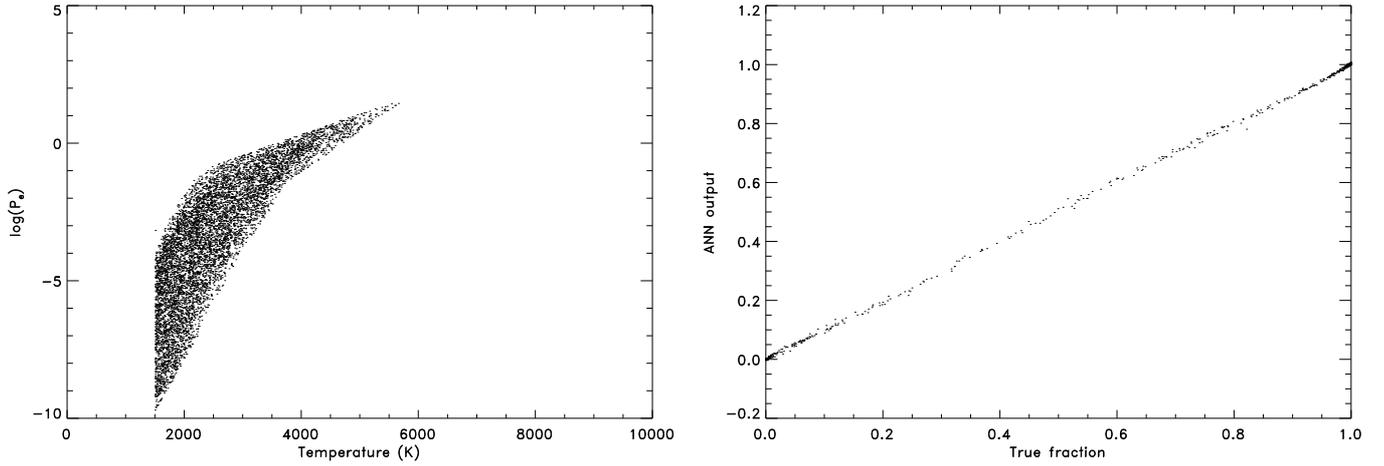}
  \caption{Left: Points in our initial training set that exhibit a
    number fraction of molecular H greater than 0.1 and smaller than
    0.9. Right: Scatter plot showing the accuracy of the ANN trained
    to retrieve the fraction of atomic H from ($T$, $P_e$ and
    $m$). The standard deviation is $\sim$5$\times$10$^{-3}$. }
  \label{fig:fracH}%
\end{figure*}

Not surprisingly, the fraction of molecular H does not cover the
entire parameter space in a uniform manner. Instead, it is saturated
in large regions of the space and there is only a
relatively narrow range of input values in which we actually need to
perform the calculation. This can be seen in Fig~\ref{fig:fracH}
(left), which shows the ($T$, $P_e$) space spanned by the training
set. The populated region has values for which the fraction of
molecular H is non-trivial. The empty space to the right is too hot
for molecules to form and therefore all H is in atomic form. To the
left, there are no ($T$, $P_e$) values consistent with our ($T$,
$P_g$) distribution.

The results shown in Fig~\ref{fig:fracH} (left panel) suggest that we
do not need to train the ANN to operate in the full domain of the
input parameters. We need to cover only the populated region seen in
the figure. In this manner we not only decrease the required size of
the training set but also improve the accuracy of a given ANN since it
can become more specialized by operating on a smaller subspace. We
therefore constructed a new, more optimal, training set that includes
only points in the relevant range. After successfully training the ANN
with these points, we reached an accuracy (measured as the standard
deviation of the difference between the validation set and the ANN
result) of $\sim$5$\times$10$^{-3}$ (see right panel of
Fig~\ref{fig:fracH}). Our ANN has four non-linear layers with 10
neurons per layer.

Once we know how much H is in the form of molecules, we use the Saha
ionization equation to compute all the relevant populations. We stress
that, even though this method only gives us the abundance of the H
molecule, it has been computed taking into account all others with the
full ICE procedure. 

For the thermodynamical variables we have the following options:

\subsection{The Wittman procedure}

The first option is the method of \citet{W74}, which in turn is an
improvement over the one introduced by \citet{M67}. It is the method
implemented in the SIR code of \citet{RCdTI92}. Only H
molecules are considered here, thus removing the necessity for
iterations and speeding up the computation of the total gas pressure
$P_g$. With this procedure $P_g$ is obtained directly from the pair of
values ($T$, $P_e$). The reverse process, i.e.  obtaining $P_e$ from
($T$, $P_g$), requires iteration from an initial guess, which is
slower and could potentially fail to converge. This method is a good
approximation in most conditions except for very cool plasmas, such as
those in a sunspot umbra, where other molecules might be important.

\subsection{Artificial Neural Networks}

\begin{figure}
  \centering
  \includegraphics[width=0.5\textwidth]{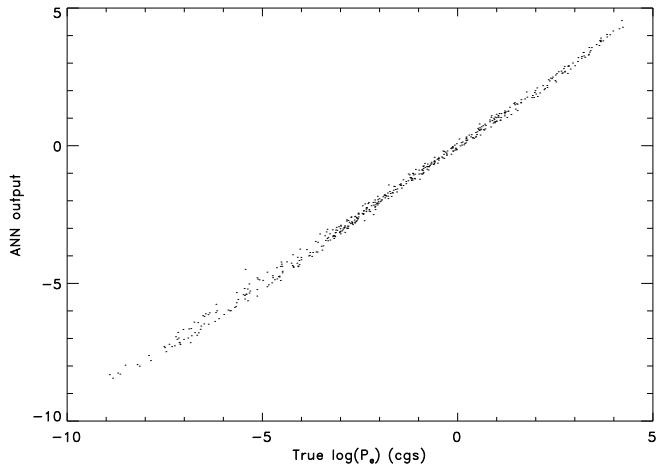}
  \caption{Scatter plot showing the accuracy of the ANN trained to
    retrieve the logarithm of $P_e$ from from ($T$, $P_g$ and
    $m$). The standard deviation is $\sim$15\%. }
  \label{fig:ANN_pepg}%
\end{figure}

It is possible to train a set of ANNs to solve for both $P_e$ from
($T$, $P_g$) and $P_g$ from ($T$, $P_e$). However, this calculation is
far less accurate than the calculation of the H molecular fraction
explained above (at least using a similar sized
ANN). Figure~\ref{fig:ANN_pepg} shows the spread in the validation
set. The error in the logarithm of the retrieved pressure is of the
order of 15\%. On the other hand, there are many applications in which
an accurate solution of the EoS is not required, since spectral lines
are far less sensitive to density or gas pressure than they are to
temperature. If the penalty in accuracy is acceptable then this method
is by far the fastest and provides a direct solution in both
directions.

\subsection{NICOLE EoS}

This is essentially the same procedure as the NICOLE method described
above, using an ANN to determine the fraction of molecular H, but then
solving the Saha ionization equation for the rest of atomic species to
determine the electron number densities. The reverse process,
i.e. obtaining $P_g$ from ($T$, $P_e$) is done by iteration, just as
in the Wittmann procedure.

\section{Background opacities}

Background opacities are those resulting from continuum absorption
processes, typically atomic photoionization. We distinguish two
distinct wavelength regimes that are treated differently: ultraviolet
and visible/infrared. The transition between those two regimes is
located at 400~nm.

\subsection{Visible and infrared opacities}

NICOLE contains three different opacity packages for the calculation
of background opacities in the visible and infrared. They
account for almost the same physical processes (with the slight
differences that we detail below) and therefore differ mostly in
details such as the tabulated values employed or the actual coding.

\subsubsection{The Wittmann package}

This package computes continuum opacities due to H$^-$, neutral H,
He$^-$, H$_2^-$, H$_2^+$, photoionization of Ca, Na and Mg, and
Rayleigh scattering by neutral H, H$_2$, neutral He and Thomson
scattering by electrons. For more details see \citet{W74}.

\subsubsection{The SOPA package}

We implemented a module with the background opacity package of
\citet{KSR96}, which includes neutral H, H$^-$, H$_2^+$,
photoionizations from the first 8 levels of Si, C, Mg, Al, and the
first two levels of Fe, Rayleigh scattering by neutral H and Thomson
scattering by free electrons. Unfortunately we were not able to bring
this package to the coding standards of the NICOLE requirements. In
order to avoid compile problems or hardware incompatibilities this
package is not supported. It is disabled by default and available only
via a special compilation-time switch for advanced users.

\subsubsection{The NICOLE package}

We developed an independent opacity package for NICOLE that computes
opacities from neutral H, H$^-$ and Mg, as well as scattering due to
H, H$_2$ and free electrons.

\begin{figure}
  \centering
  \includegraphics[width=0.5\textwidth]{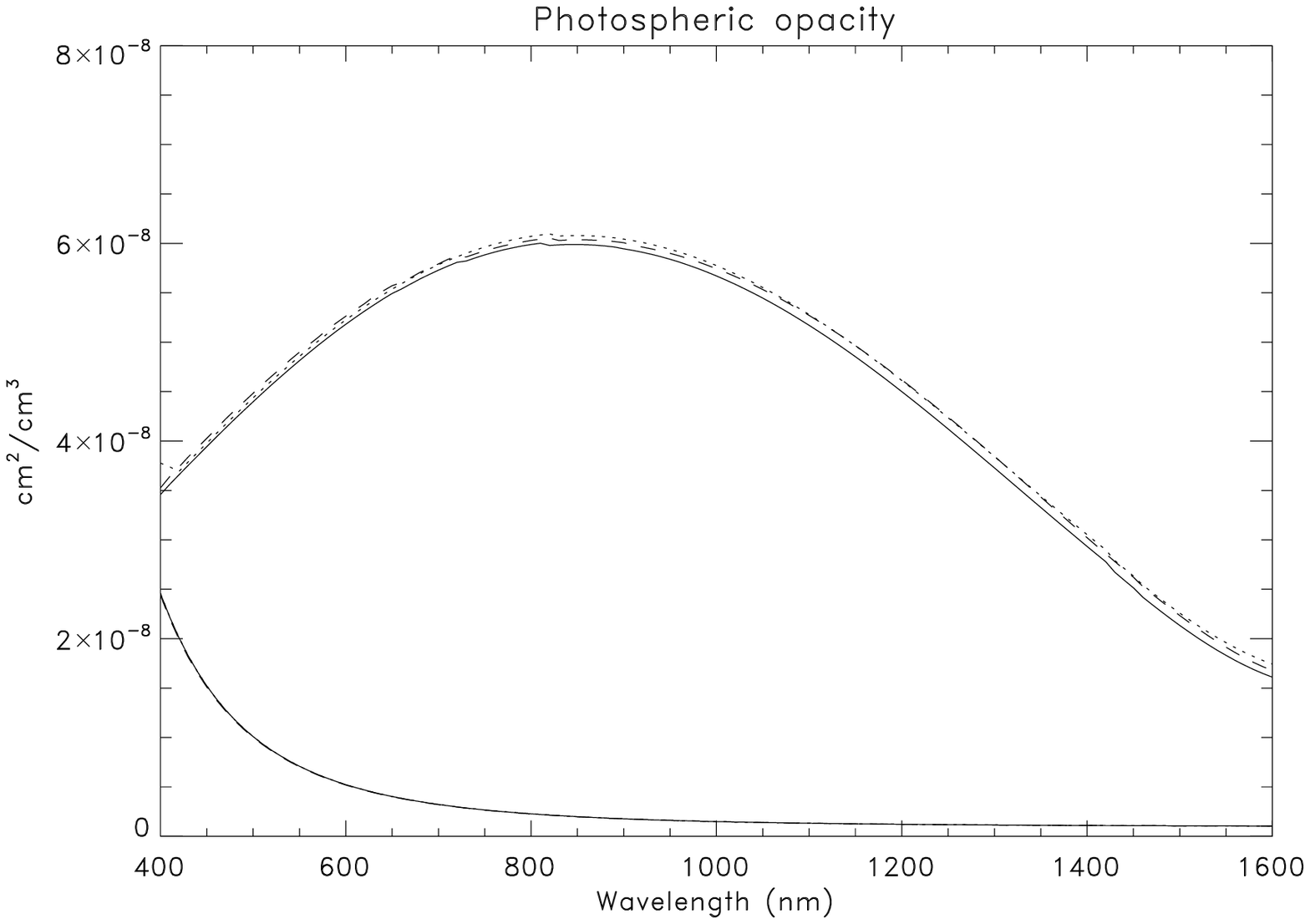}
  \caption{Background opacities as a function of wavelength in
    typically photospheric conditions ($T$=5600~K,
    $P_e$=10~dyn~cm$^{-2}$,
    $P_g$=8$\times$10$^5$~dyn~cm$^{-2}$). Solid line: Using the
    Wittmann package. Dashed line: Using the NICOLE package. Dotted
    line: Using the SOPA package. The lower curve represents the
    scattering contribution to the opacity (all three packages yield
    the same result within the line thickness of the plot). The
    scattering curve has been multiplied by a factor 100 for better
    visibility in this plot.}
  \label{fig:opac_ph}%
\end{figure}

\begin{figure}
  \centering
  \includegraphics[width=0.5\textwidth]{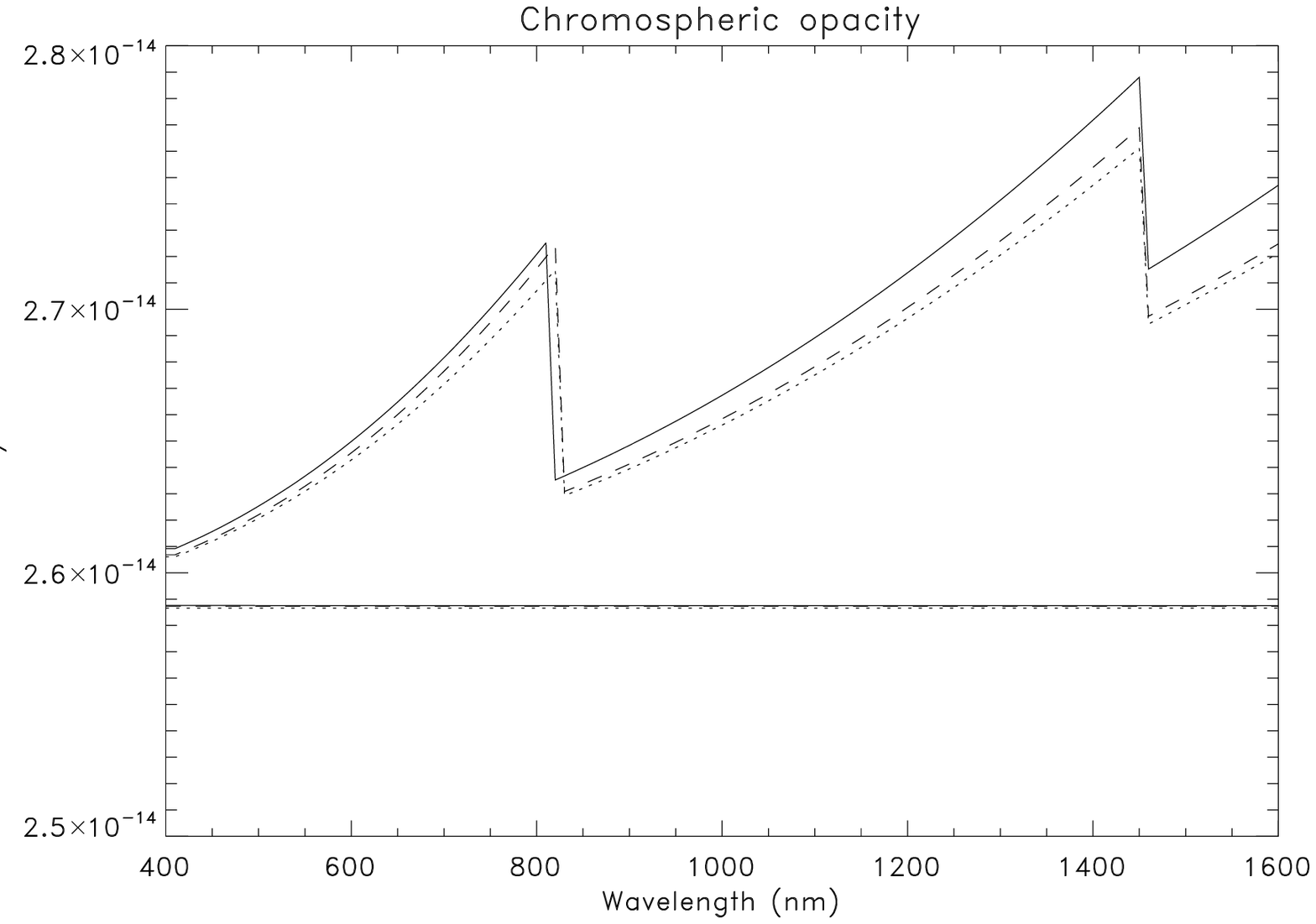}
  \caption{Background opacities as a function of wavelength in
    typically chromospheric conditions ($T$=9000~K, $P_e$=0.05~dyn~cm$^{-2}$, $P_g$=0.15~dyn~cm$^{-2}$). Solid line:
  Using the Wittmann package. Dashed line: Using the NICOLE
  package. Dotted line: Using the SOPA package. The lower curve
  represents the scattering contribution to the opacity (all three
  packages yield the same result within the line thickness of the
  plot).}
  \label{fig:opac_ch}%
\end{figure}

Figures~\ref{fig:opac_ph} and~\ref{fig:opac_ch} compare the wavelength
dependence of the background opacities computed by all three
packages. Since this comparison depends strongly on the atmospheric
conditions, we have chosen two sets of parameters.

\begin{figure}
  \centering
  \includegraphics[width=0.5\textwidth]{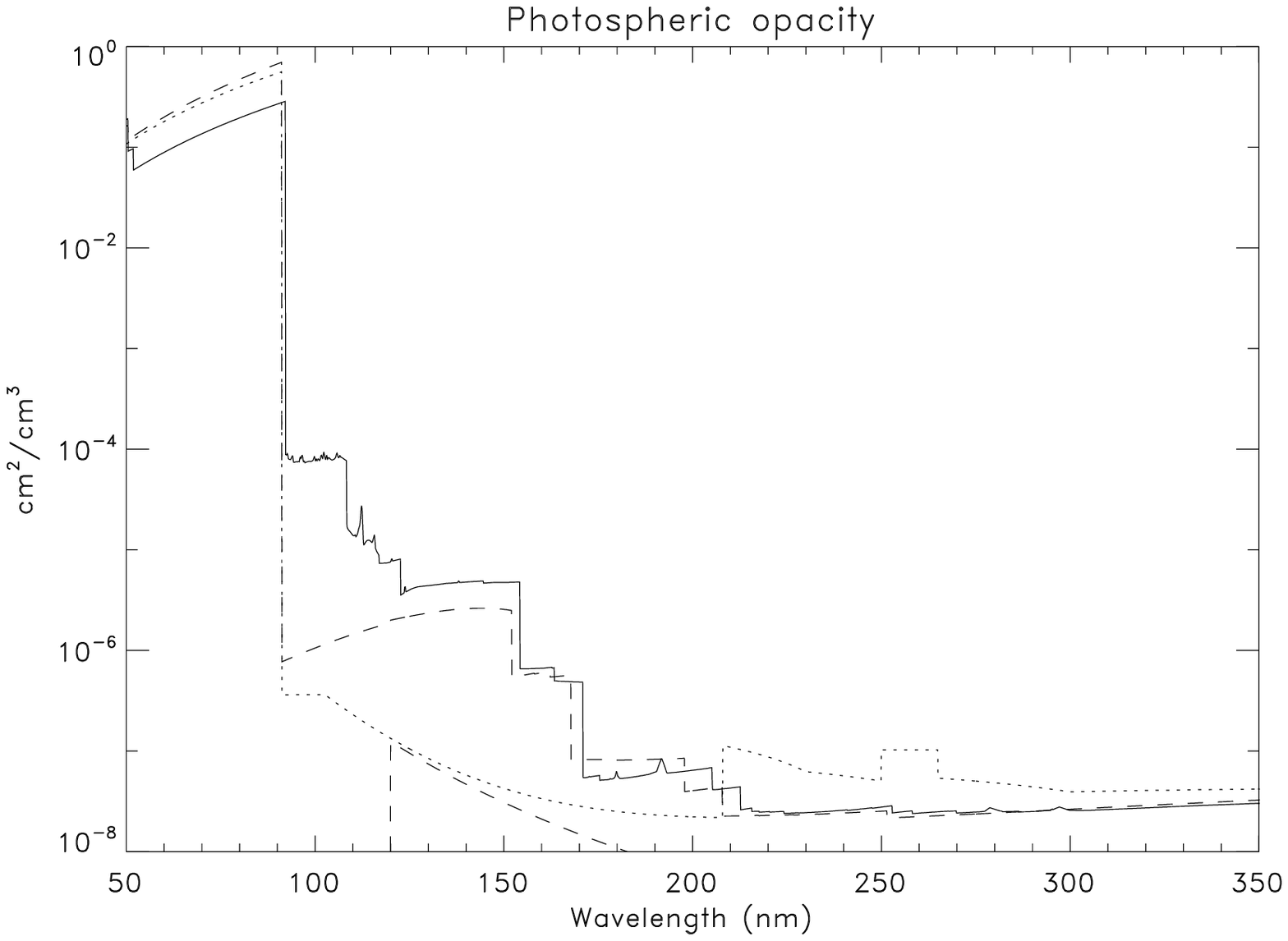}
  \caption{Ultraviolet background opacities as a function of
    wavelength in typically photospheric conditions ($T$=5600~K,
    $P_e$=10~dyn~cm$^{-2}$,
    $P_g$=8$\times$10$^5$~dyn~cm$^{-2}$). Solid line: Using the
    TIP-TOP package. Dashed line: Using the Dragon-Mutschlecner
    package. Dotted line: Using the SOPA package. The lower dashed curve
    represents the scattering contribution to the opacity (all three
    packages yield the same result within the line thickness of the
    plot). }
  \label{fig:opac_ph_uv}%
\end{figure}

\begin{figure}
  \centering
  \includegraphics[width=0.5\textwidth]{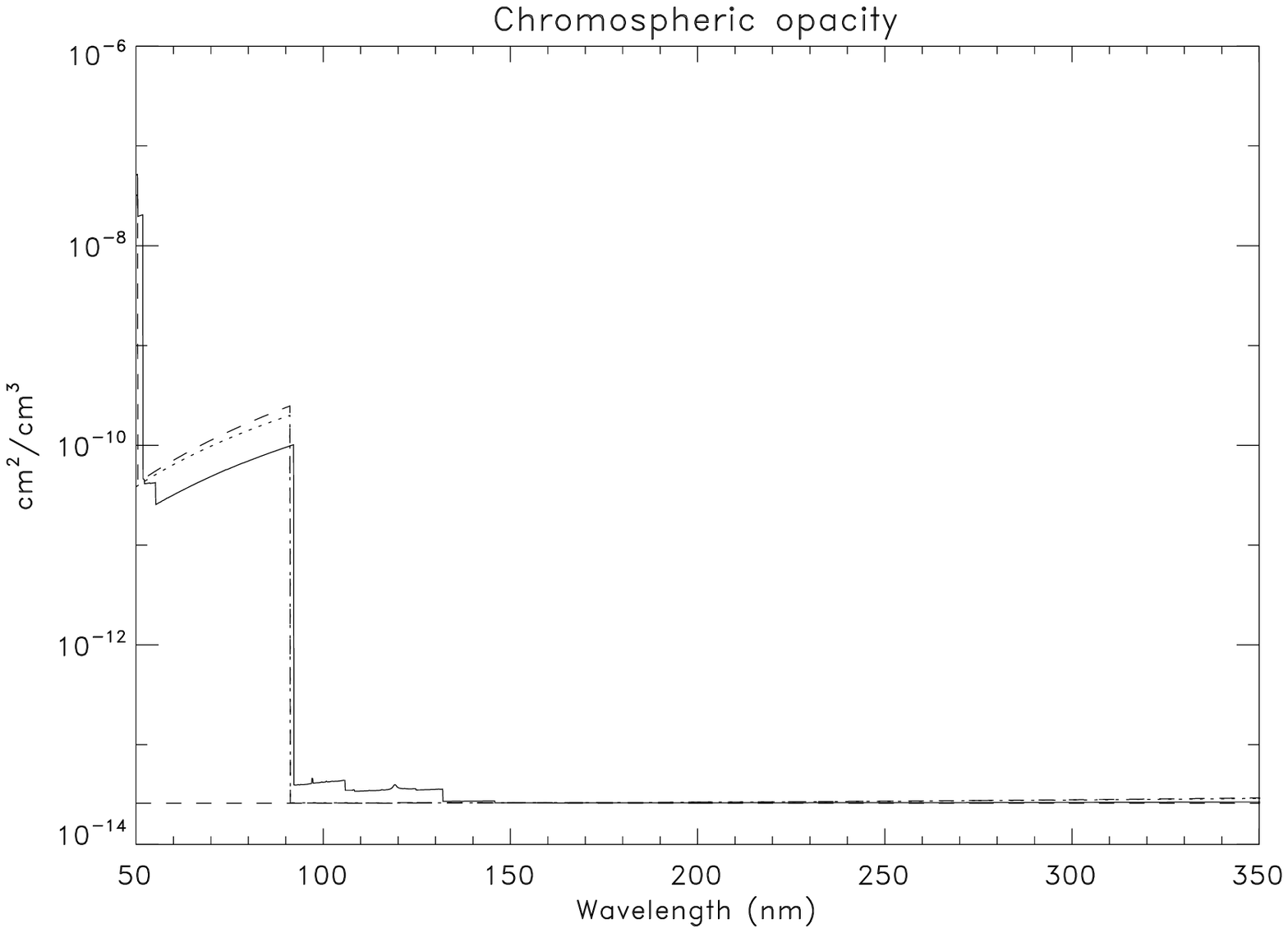}
  \caption{Ultraviolet background opacities as a function of
    wavelength in typically chromospheric conditions ($T$=9000~K,
    $P_e$=0.05~dyn~cm$^{-2}$, $P_g$=0.15~dyn~cm$^{-2}$). Solid line:
    Using the TIP-TOP package. Dashed line: Using the
    Dragon-Mutschlecner package. Dotted line: Using the SOPA
    package. The lower dashed curve represents the scattering contribution to
    the opacity (all three packages yield the same result within the
    line thickness of the plot).}
  \label{fig:opac_ch_uv}%
\end{figure}

\begin{figure*}
  \centering
  \includegraphics[]{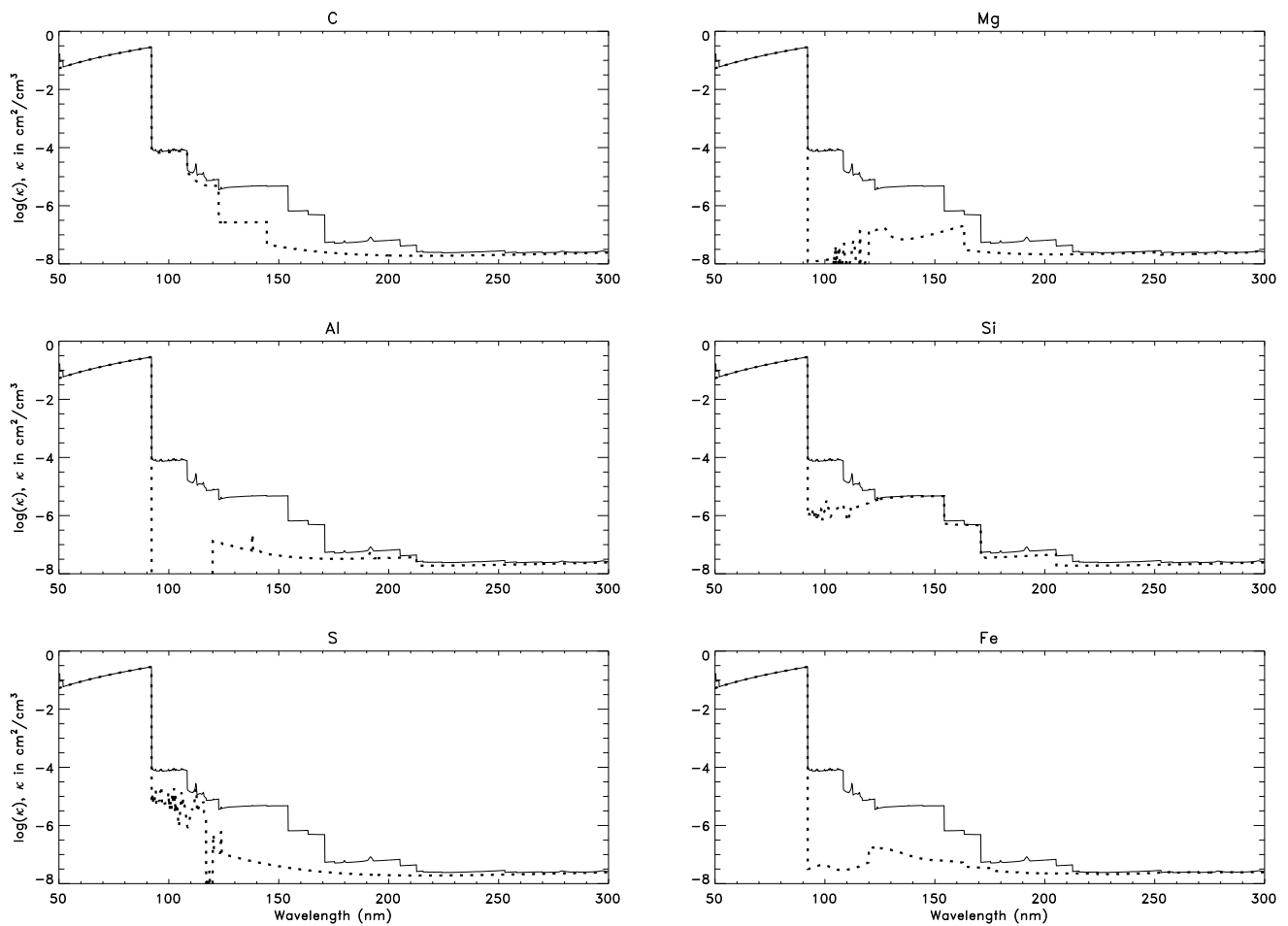}
  \caption{Contribution of the most relevant metals to the background
    ultraviolet opacity in typical photospheric conditions
    ($T$=5600~K, $P_e$=10~dyn~cm$^{-2}$,
    $P_g$=8$\times$10$^5$~dyn~cm$^{-2}$). Other elements become
    relevant under different conditions. Solid: Total opacity. Dotted:
    Resulting opacity when neglecting all metals except for the one
    indicated in the title of each plot.}
  \label{fig:opac_uv_elem}%
\end{figure*}

\subsection{Ultraviolet opacities}

Computation of background opacities in the ultraviolet is far more
complicated than in the visible. Many metallic species can undergo
photoionization processes with sufficiently large cross-sections to
become important opacity contributors in spite of their relatively low
abundances. To complicate the matter further, there is not a dominant
species above the 91~nm regime, where the H photoionization
occurs. Depending on the prevailing conditions and the wavelength
range, we are dominated by different metals. 
In addition to the SOPA package (see
above), which includes some photoionization processes for a few
interesting metals, we have two other packages in NICOLE specifically
implemented for the computation of ultraviolet opacities.

\subsubsection{The Dragon-Mutschlecner package}

\citet{DM80} provide a set of tables to compute photoionization
cross-sections for various levels of neutral Mg, Al, Si and Fe. Using
some simple analytical expressions, we can obtain a rather good
approximation in most practical situations (see
Figures~\ref{fig:opac_ph_uv} and~\ref{fig:opac_ch_uv} below).

\subsubsection{The TIP-TOP package}

The Iron Project (TIP) and The Opacity Project (TOP) are two large
collaborations aimed at producing the most comprehensive compilation
of atomic opacity sources. The two projects started off as independent
initiatives but have now joined forces and have published their tables
with a large number of photoionization cross-sections for most
metals. We have included all the available data for neutral and singly
ionized elements between $Z$=1 and $Z$=26. In the particular case of
the Fe atom, we use the data provided by~\citet{B97} and
by~\citet{NP94}. In all cases the data are smoothed as discussed in
\citet{BRP98,APLH+03,AP08}.

To make the problem tractable, NICOLE preloads in memory a large
matrix with all the cross-sections (for each element and level) at
each wavelength, discretized with a 0.1~nm sampling. When the opacity
routine is called for a certain wavelength and input conditions, it
simply picks from the matrix all the cross-sections at that wavelength
(rounded off to the closest point in the grid), weighs each one
according to element abundance, ionization fraction and level
excitation, and finally returns the total of all the
contributors. With this strategy, we can obtain the total opacity from
all contributors in the comprehensive TIP-TOP database in a very short
time.

Figures~\ref{fig:opac_ph_uv} and~\ref{fig:opac_ch_uv} compare the
various ultraviolet opacity packages in two different situations. In
the photosphere (Fig~\ref{fig:opac_ph_uv}), the TIP-TOP package yields
a much more detailed curve with a plethora of peaks and
discontinuities caused by photoionization from countless levels of
several elements. The other two packages, however, produce a good
smoothed out approximation. Under these atmospheric conditions, the
opacity is dominated by neutral species. Under the chromospheric
conditions of Fig~\ref{fig:opac_ch_uv}, the opacity structure is much
simpler. It is dominated by Thomson scattering on free electrons at
all but the shortest wavelengths. Under these conditions we start to
find some non-negligible contributions from ionized metals.

\section{Formal solutions}

Inside each vertical column, NICOLE solves the NLTE problem in 1D by
assuming plane-parallel geometry, isotropic scattering and complete
frequency redistribution \citep[details in][]{SNTB97}. To compute the
atom population densities, it assumes statistical equilibrium and
unpolarized light. Once the populations are known, the full-Stokes
vector is computed for Zeeman-induced polarization. Therefore, all
Zeeman sublevels originating from a given atomic level are assumed to
be equally populated, discarding any quantum interference between them
\citep[the \emph{polarization-free}
  approximation,][]{1996SoPh..164..135T}.

\subsection{Formal solutions and the NLTE problem}
\label{sec:formalsolvers}

The radiative transfer equation for unpolarized light can be expressed as:
\begin{equation}
  \frac{dI_\nu}{d\tau_\nu} = I_\nu - S_\nu, \label{eq:rad}
\end{equation}
where $I_\nu$ is the emerging intensity at frequency $\nu$, $\tau_\nu$
is the optical depth and $S_\nu$ is the source function. In a discrete
grid of depth points where the subindexes $u$, $o$ and $d$ indicate
the upwind point, central point and downwind point respectively, the solution to Eq.~\ref{eq:rad} on the interval ($\tau_o, \tau_u$) is:
\begin{equation}
I(\tau_o) = I(\tau_u) e^{-\delta_o} + \int^{\tau_u}_{\tau_o}
e^{-(\tau-\tau_o)} S(\tau)d\tau,\label{eq:sol}
\end{equation}
where $\delta_o \equiv \delta\tau_o = |\tau_u - \tau_o|$. To integrate analytically Eq.~(\ref{eq:sol}), the source function can be approximated using a polynomial interpolant: linear, quadratic, etc. We have implemented two formal solutions of the radiative transfer equation (methods) to compute the atom population densities (unpolarized), based on short-characteristics:
\begin{enumerate}
  \item The source function is approximated with a parabolic interpolant, centered in the grid point where the intensity is being calculated \citep{1987JQSRT..38..325O}. This interpolant behaves particularly well on equidistant grids, but it is known to overshoot in irregular grids. Therefore, when overshooting is detected, we adopt a linear approximation instead. \label{it:sc}
    \item An elegant approach, introduced by \citet{2003ASPC..288....3A}, is to use Bezier-splines interpolants. Bezier-splines provide a powerful framework to control overshooting, while keeping the accuracy of high-order interpolants. These methods have been implemented in 3D MHD codes \citep[e.g., BIFROST,][]{2010A&A...517A..49H} and radiative transfer codes \citep[e.g., Multi3d, PORTA, ][]{2009ASPC..415...87L,STB13}.\label{it:bez}
\end{enumerate}

In particular, we have experimented extensively with the election of an appropriate diagonal approximate lambda operator and the treatment of overshooting cases in Method~\ref{it:bez}. The quadratic Bezier interpolant is defined using normalized abscissa units in the interval ($x_o,x_{u}$):
\begin{equation*}
  u = \frac{x-x_{u}}{x_o - x_{u}},
\end{equation*}
so,
\begin{equation}
 f(x) = y_{u} u^2 + y_o(1-u)^2 + 2u(1-u)\cdot C,\label{eq:bezierspline}
\end{equation}
where C is a control point defined as:
\begin{equation}
  C = S_o - \frac{\delta_o}{2} \frac{dS_0}{d\tau}.\label{eq:control}
\end{equation}

The solution to Eq.~\ref{eq:sol}, can be formulated in two ways. One
is to re-arrange the terms of the integral so we get terms that only
depend on the values of the source function in the upwind point ($u$),
downwind point ($d$) and central point ($o$), by explicitly replacing the
Eq.~\ref{eq:control} into Eq.~\ref{eq:bezierspline} before the
integral in Eq.~\ref{eq:sol} is performed:
\begin{equation}
  I_o = I_{u}e^{-\delta_o} + \alpha S_{u} + \beta S_o + \gamma S_{d},\label{eq:bez1}
\end{equation}
where $\alpha, \beta, \gamma$ are the interpolation coefficients. This
is the choice of \citet{2010A&A...517A..49H}, therefore they choose an
expresion for computing numerically $dS_0/d\tau$ that linearly depends
on $S_{u},S_o,S_{d}$. However, \citet{2013ApJ...764...33D} and
\citet{STB13} express the solution as a function of
$S_{u},S_o,C$:
\begin{equation}
  I_o = I_{u}e^{-\delta_o} + \hat{\alpha} S_{u} + \hat{\beta} S_o +
  \hat{\gamma} C. \label{eq:bez2}
\end{equation}

In principle, both formalisms should be equivalent. The only
difference appears when computing the diagonal approximate lambda
operator \citep[details in][]{1987JQSRT..38..325O}. To define the
approximate lambda operator, one can use a source function that is set
to zero at all depth points except $S_o=1$, and check what terms
remain in Equation~\ref{eq:bez1} and \ref{eq:bez2}. Note that we are
strictly neglecting the contribution from the ensuing intensity through the term
$I_{u}e^{-\delta_o}$, which is typically very small in the optically
thick regime (\citealt{2002ApJ...568.1066V}).

The implementation
by \citet{2010A&A...517A..49H} implicitly includes the terms used to
compute $dS_0/d\tau$ in their operator, whereas the second formalisms
does not. It is straightforward to see that those terms appear because
the derivative is computed numerically, but there is no reason to
include them in the approximate operator. Also, by using the second
formalism, it does not matter so much what expresion is used for the
derivative, given that the control point is not explicitely split into
terms that depend on $S_{d}$, $S_o$ and $S_{u}$
\citep[][proposes two different ways of
computing centered derivatives]{2003ASPC..288....3A}. Our tests show
that defining the local operator as
$\Lambda^*=\hat{\beta}+\hat{\gamma}$ is optimal and convergence is
achieved in less iterations. 

Overshooting is suppressed by changing the value of the control point $C$ as described in \citet{2010A&A...517A..49H}. The basic idea is to identify extrema in the source function, and to constrain the value of the control point within $S_{u}$ and $S_o$. An important refinement is proposed by \citet{STB13}, who also check for overshooting in the downwind interval, between $S_{o}$ and $S_{d}$. The latter indeed improves the stability of the solution, forcing the Bezier interpolant to approach point $o$ monotonically in every situation.

So far, we have not mentioned much about Method~\ref{it:sc}. The problem is that allowing the solution to switch from parabolic to linear and vice-versa can lead to a flip-flop behaviour. Normally, more iterations are needed to reach a similar convergence threshold with this method than with the Bezier alternative.

\subsection{Formal solution of the polarized transfer equations}
Once the level populations are calculated, NICOLE allows to compute the emerging Stokes vector assuming Zeeman-induced polarization. We have implemented a list of formal solvers that can be used for this matter. The following alternatives are available:
\begin{itemize}
  \item Quadratic and cubic DELO-Bezier \citep[][]{2013ApJ...764...33D}.
    \item DELO-Linear \citep[][]{1989ApJ...339.1093R}.
      \item DELO-Parabolic \citep[][]{2003ASPC..288..551T}.
        \item Hermitian \citep{1998ApJ...506..805B}.
          \item Weakly-polarizing media \citep{1999ApJ...526.1013S}.
\end{itemize}

\section{Hyperfine structure}
Almost every element in the periodic table has an isotope with
non-zero nuclear angular momentum $I$, which couples with the sum of
the orbital and spin angular momentum $J$. Consequently, the fine
structure levels, characterized by their value of $J$, are split into
hyperfine structure levels following the standard rule for angular
momentum addition, yielding $F=|J-I| \ldots J+I$.  The hyperfine
splitting is usually much smaller than the fine structure
splitting. Thus, the presence of a weak magnetic field may be able to
produce Zeeman splittings that are of the order of the energy level
separation between consecutive $F$ levels. Under these circumstances,
the non-diagonal terms in the Zeeman Hamiltonian become of
importance. This regime of intermediate Paschen-Back effect (or
Back-Goudsmit effect) leads to strong perturbations on the Zeeman
patterns, which may have an important impact on the emergent Stokes
profiles.

The energy splitting of these $F$ levels with respect to the original
$J$ level (the fine structure level without hyperfine structure) is
given, with very good approximation, by \citet{casimir63}:
\begin{eqnarray}
\Delta_\mathrm{HFS}(J,F,I) &=& \frac{1}{2} AK \nonumber \\
&+& \frac{1}{2} B \frac{(3/4)K(K+1)-I(I+1)J(J+1)}{I(2I-1)J(2J-1)},
\end{eqnarray}
where
\begin{equation}
K=F(F+1)-I(I+1)-J(J+1).
\end{equation}
The energy splitting is represented in cm$^{-1}$ when the constants
$A$ and $B$ are given in cm$^{-1}$. These constants are the
magnetic-dipole ($A$) and electric-quadrupole ($B$) hyperfine
structure constants, and are characteristic of a given fine structure
level. In the case that the energy level separation between
consecutive fine structure levels is very large in comparison to the
typical Zeeman splitting produced by the magnetic fields we are
interested in, one may focus exclusively on the coupling between the
hyperfine and magnetic interactions. The total Hamiltonian is
block-diagonal and each block can be written as
\citep[e.g.,][]{landi_landolfi04}:
\begin{align}
\langle &(LS)J I F M_F | H | (LS) J I F' M_F' \rangle
= \delta_{FF'} \delta_{M_F M_F'} \Delta_\mathrm{HFS}(J,F,I) \nonumber \\
&+ \delta_{M_F M_F'} \mu_0 B g_J (-1)^{J+I-M_F} \nonumber 
\sqrt{J(J+1)(2J+1)(2F+1)(2F'+1)} \nonumber \\
&\times \sixj{F'}{F}{1}{J}{J}{I} \threej{F}{F'}{1}{-M_F}{M_F}{0},
\end{align}
where $\mu_0$ is the Bohr magneton, $B$ is the magnetic field strength and $g_J$ is the 
Land\'e factor of the level in L-S coupling.

The total Hamiltonian is diagonal in $M_F$, so that it remains a good
quantum number even in the presence of a magnetic field. This is not
the case with $F$, 
because the total Hamiltonian mixes levels with different values of
$F$. After a numerical diagonalization of the Hamiltonian, the
eigenvalues are associated with the energies of the $M_F$ magnetic
sublevels. The transition between the upper and lower fine structure
levels produce many allowed transitions following the selection rules
$\Delta M_F=0,\pm 1$. The strength of each component can be obtained
by evaluating the squared matrix element of the electric dipole
operator \citep{landi_landolfi04}:
\begin{equation}
S_q^{i M_F, i' M_F'} \propto | \langle (LS) J I i M_F | r_q | (LS) J I i' M_F' \rangle |^2,
\end{equation}
where $q=M_F-M_F'=0,\pm 1$ and $| (LS) J I i M_F \rangle$ are the eigenvectors of the Hamiltonian. 
The symbol $i$ is used for identification purposes since $F$ is not a good quantum number 
\citep[e.g.,][]{landi_landolfi04}.

\section{Abundance inversions}

When working in inversion mode, it is possible to set element
abundances as ``inversion nodes''. In this manner, NICOLE can be used
in studies of solar and stellar chemical compositions, similarly to
the MISS code of \citet{APBA+01}. In principle it is
possible to invert abundances and atmospheric parameters
simultaneously. However, it is important to realize that doing this
would only produce meaningful results if the observations include
lines from multiple elements that can univocally constrain both the
atmosphere and the composition. In general, it is better to have
independent observations to determine the atmospheric model, or at
least to have a good approximation to it before attempting to invert
abundances.

\begin{figure}
  \centering
  \includegraphics[width=0.5\textwidth]{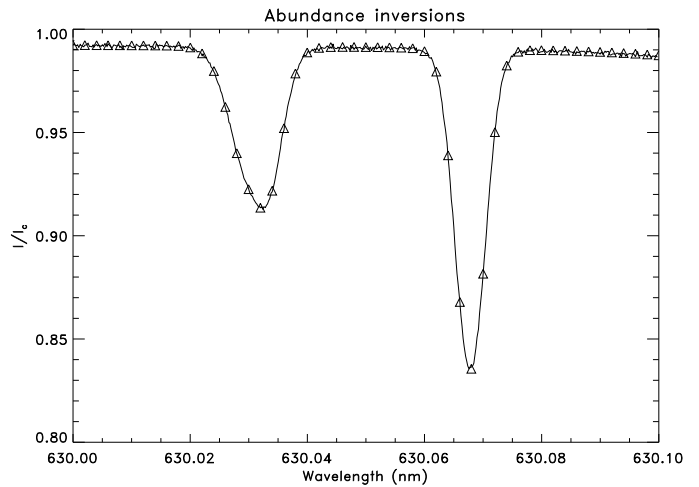}
  \caption{Simulated inversion of three element abundances: O, Ni and Sc}
  \label{fig:abundinv}%
\end{figure}

Figure~\ref{fig:abundinv} shows a simulated inversion of the
well-known blend of \NiI \, with a forbidden \OI \, transition at
6300.3~\AA \, along with the nearby \ScII \, line. These lines have
been frequently used in recent studies of the solar chemical
composition, as this region has proven to be a valid diagnostics to
resolve the so-called solar Oxygen crisis (e.g., \citealt{SN14} and
references therein). The simulated observations were synthesized with
the HSRA quiet Sun model \citep{GNK+71}, adding random noise of a
1-$\sigma$ amplitude of 5$\times$10$^{-4}$. The reference abundances
chosen in this test for the lines in the figure are 8.83, 6.25 and
3.17 (O, Ni and Sc, respectively). The inversion was initialized with
highly discrepant values: 8.00, 5.50 and 4.00 and repeated up to 30
times adding a random perturbation of up to $\pm$0.2 dex to the
reference values. Only 8 out of the 30 inversions converged to the
correct solution, producing a fit down to the noise level. The fit
shown in the figure is representative of these 8 solutions, having a
$\chi^2$ value that is approximately that of their average. The mean
value and standard deviation of the results from the inversions are
8.835$\pm$0.004, 6.254$\pm$0.004 and 3.174$\pm$0.004, respectively.

It is important to note that these extremelly small uncertainties
represent only the inversion error. In this case the model atmosphere
is prescribed and known {\it a priori} because we are interested here
in the error produced by the inversion process and the algorithm's
ability to find the correct solution. Otherwise, one would also have
systematic errors arising from the atmospheric model uncertainty,
which are likely much larger.

\section{Parallelization}

NICOLE was designed to work on large datasets, typically inversions of
spectral (or, in general, spectro-polarimetric) scans of a 2D field of view
\citep{2013A&A...556A.115D} or spectral synthesis in simulation datacubes
\citep{SN14}. For such applications, an efficient parallelization
scheme is required. We have implemented a master-slave approach in
which each slave works on a given spatial pixel. All input and output
tasks are handled by the master process, which reads the input files,
sends the input data to each idle slave, collects the computation
results and finally writes them to disk. This strategy eliminates
possible disk access conflicts or bottlenecks among processes and is
optimal in minimizing the computing time for large problems. We
achieve the goal of ideal parallelization, in which the CPU time is
inversely proportional to the number of processors (as long as the
computation time is much longer than the time it takes to read the
input data). This ideal parallelization holds independently of the
number of processors, making NICOLE a massively parallel code which
could potentially run efficiently even on the largest supercomputers
with thousands of processors. We present some tests below demonstrating
this good behavior up to 200 processors.

We conducted a series of tests with a benchmark calculation using the
3D MHD model computed with the BIFROST code (see
\citealt{GCH+11}). The snapshot used in our calculations is publicly
available as part of the LMSAL IRIS mission data (\citealt{dPTL+14}) and
has been previously used in a number of studies (e.g.,
\citealt{LCRvdV12, dlCRdPC+13,PLdP+13}). The simulation is computed on
a grid of ($x$,$y$,$z$)=504$\times$505$\times$496 points,
corresponding to a physical size of approximately 24$\times$24~Mm in
the ($x$,$y$)-plane. Vertically, the simulation extends from 2.2~Mm
below the photosphere, to 15~Mm above, and it encloses a photosphere,
chromosphere and corona. We have only considered every 4th pixel in
the horizontal ($x$,$y$)-plane to be able to perform the calculation
with a reduced number of CPUs in our tests.

\begin{figure}
  \centering
  \includegraphics[width=0.5\textwidth]{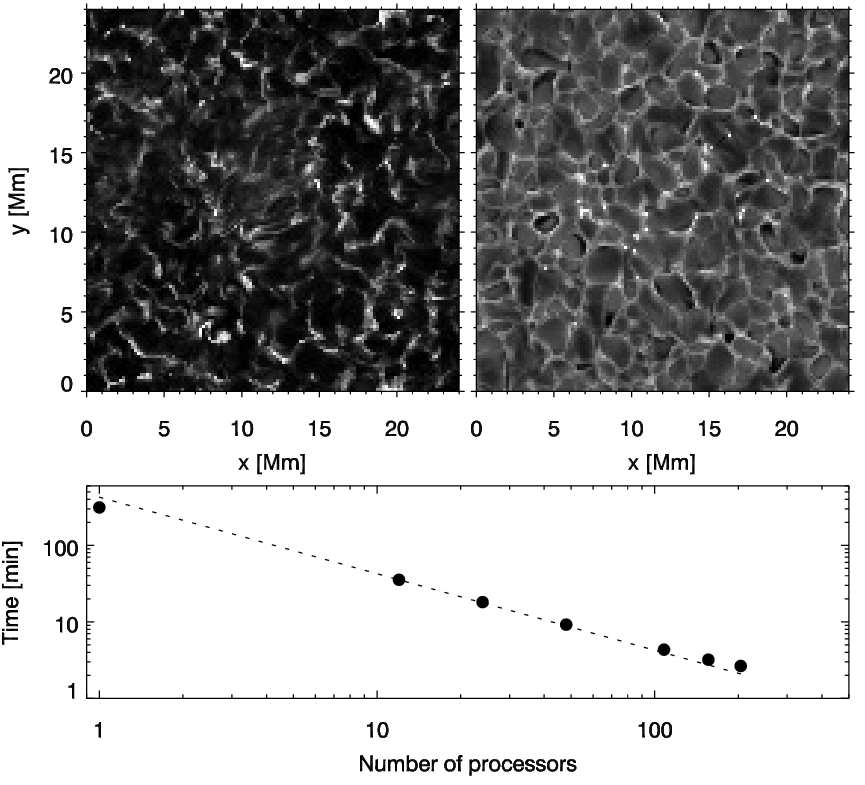}
  \caption{\emph{Top row:} Synthetic observations in the 854.2~nm
    line, close to line center (left panel) and in the extended
    photospheric wing (right panel). \emph{Bottom:} CPU time as a function of the inverse number of
    processors. The dotted line represents the behavior expected for
    ideal parallelization.  }
  \label{fig:paral}%
\end{figure}

For each column in the simulation, we solved the NLTE problem with a
6-level \CaII\ atom and computed intensity and polarization profiles in
the 8542~\AA \ line. The hardware platform is a Linux AMD Opteron
cluster with 524 cores. We employed a homogeneous subset of 204 of
them in our tests (this cluster has several different processor
models). Fig~\ref{fig:paral} shows a log-log plot of the total CPU
time versus the inverse of the number of (slave) processors
employed. In the ideal case of optimal paralellization one would
expect a straight line whose slope is $-1$ and the abscissa at origin
is the number of columns multiplied by the computing time per
column. The figure shows that the tests follow this ideal behavior
(represented in the dashed line), with no signs of saturation even at
200 processors.

Our parallelization is implemented using the MPI library. It is
straightforward to compile and run the parallel version of NICOLE on
any system with a working MPI installation. Since one of the processes
is the master, it is usually more efficient to run NICOLE with $N+1$
threads, where $N$ is the number of available hardware processor
cores.

\section{Conclusions}

NICOLE is the result of a multiyear effort to produce a public,
well-documented, user-friendly code for massive radiative transfer
calculations. It may be used in LTE or NLTE, to convert atmospheric
models between geometrical and optical depth or to make inversions of
observed profiles. It may be applied to solar or stellar models and
observations. Interested researchers are invited to download the code
from the link above and encouraged to make and redistribute any
changes or modifications they deem necessary (permissions are
explicitly granted under the GNU public license).

We expect that codes like NICOLE will become an important tool in the
coming years, at least within the solar community. The advent of new
instrumentation designed for chromospheric magnetometry will produce
enormous datasets of NLTE spectral profiles that will require
inversion. Additionally, state-of-the-art 3D numerical simulations of
the solar atmosphere, spanning the whole range from the photosphere to
the transition region, are becoming available and increasingly
realistic. Polarized spectral synthesis in the simulation datacubes
are necessary for the detailed comparison between simulations and
observations. 

NICOLE has already been tested and publications exist demonstrating
its performance in inverting Stokes profiles in LTE (\citealt{SN11}),
NLTE (\citealt{dlCRSNC+12}; \citealt{2013A&A...556A.115D};
\citealt{2014ApJ...784L..17L}) and synthesizing large numbers of
profiles in 3D atmospheric models (\citealt{SN14}).

A significant fraction of the NICOLE development effort has been
directed to making this code as user-friendly as possible. However, it
is important to remember that NICOLE, like any other complex numerical
code, cannot be used as a black box. Understanding not only the
underlying physics, but also the numerical procedures and the data
products involved, is of paramount importance in order to obtain
meaningful scientific results. 

\begin{acknowledgements}

The authors thankfully acknowledge the technical expertise and
assistance provided by the Spanish Supercomputing Network (Red
Española de Supercomputación), as well as the computer resources used:
the LaPalma Supercomputer, located at the Instituto de Astrofísica de
Canarias.

Financial support by the Spanish Ministry of Science and Innovation
through project AYA2010-18029 (Solar Magnetism and Astrophysical
Spectropolarimetry) is gratefully acknowledged. AAR also acknowledges
financial support through the Ramon y Cajal fellowship.

\end{acknowledgements}

\bibliographystyle{aa}


\end{document}